\documentclass[
    ,final 
  ]
  {aipproc}

\layoutstyle{8x11double}
\begin{document}

\newcommand{\lsim}   {\mathrel{\mathop{\kern 0pt \rlap
  {\raise.2ex\hbox{$<$}}}
  \lower.9ex\hbox{\kern-.190em $\sim$}}}
\newcommand{\gsim}   {\mathrel{\mathop{\kern 0pt \rlap
  {\raise.2ex\hbox{$>$}}}
  \lower.9ex\hbox{\kern-.190em $\sim$}}}

\title{Anti-GZK effect in UHECR spectrum}

\classification{PACS numbers: 01.30.Cc, 13.85.Tp, 98.62.Ra, 95.85.Sz}
\keywords      {cosmic rays, diffusion, intergalactic magnetic field}

\author{R. Aloisio\footnote{Talk presented by R. Aloisio} 
~~and~ V.S. Berezinsky}
{address={INFN - Laboratori Nazionali del Gran Sasso, Assergi (AQ), Italy}}

\begin{abstract}
In this paper we discuss the anti-GZK effect that arises in the framework 
of the diffusive propagation of Ultra High Energy (UHE) protons. This effect
consists in a jump-like increase of the maximum distance from which UHE 
protons can reach the observer. The position of the jump is independent of the 
Intergalactic Magnetic Field (IMF) strength and depends only on the energy 
losses of protons, namely on the transition energy from adiabatic and 
pair-production energy losses. The Ultra High Energy Cosmic Rays (UHECR) 
spectrum presents a low-energy steepening approximately at this energy, which 
is very close to the position of the observed second knee. The dip, seen in 
the universal spectrum as a signature of the proton interaction with the 
Cosmic Microwave Background (CMB) radiation, is also present in the case of 
diffusive propagation in magnetic fields.
\end{abstract}

\maketitle

\section{Introduction}
\label{introduction}

Recently a very interesting phenomenon, determined by the UHE proton 
propagation in IMF, has been found \cite{Lemoine,anti-GZK}. 
It consists in a low energy steepening of the proton spectrum that 
occurs at energy below $10^{18}$ eV. This steepening is caused by an 
increase of the diffusive propagation time, that rapidly exceeds the age 
of the universe, and can be explained trough the diffusive propagation of 
UHECR in IMF. The position of the steepening energy 
$E_{\rm s}=1\times 10^{18}$ eV,
is determined only by the proton energy losses on CMB and coincides with 
a good accuracy to the position of the 2nd knee observed in the CR 
spectrum \cite{model}. In this paper we will discuss the main features of 
the diffusive propagation of UHECR in IMF, focusing our attention on the 
steepening energy scale $E_s$. Before entering the details of the diffusive 
propagation of UHECR protons let us review the main experimental evidences 
related to IMF.

\begin{figure}[t!]
\begin{tabular}{ll}
\includegraphics[width=0.45\textwidth]{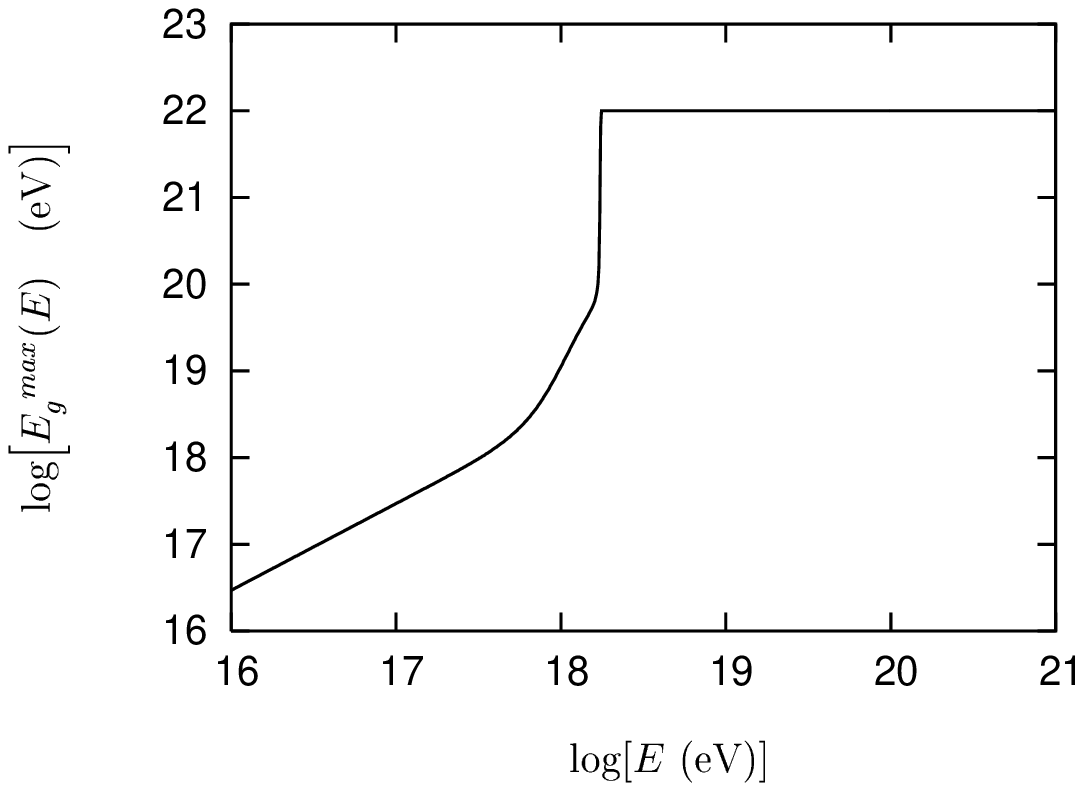}
&
\includegraphics[width=0.45\textwidth]{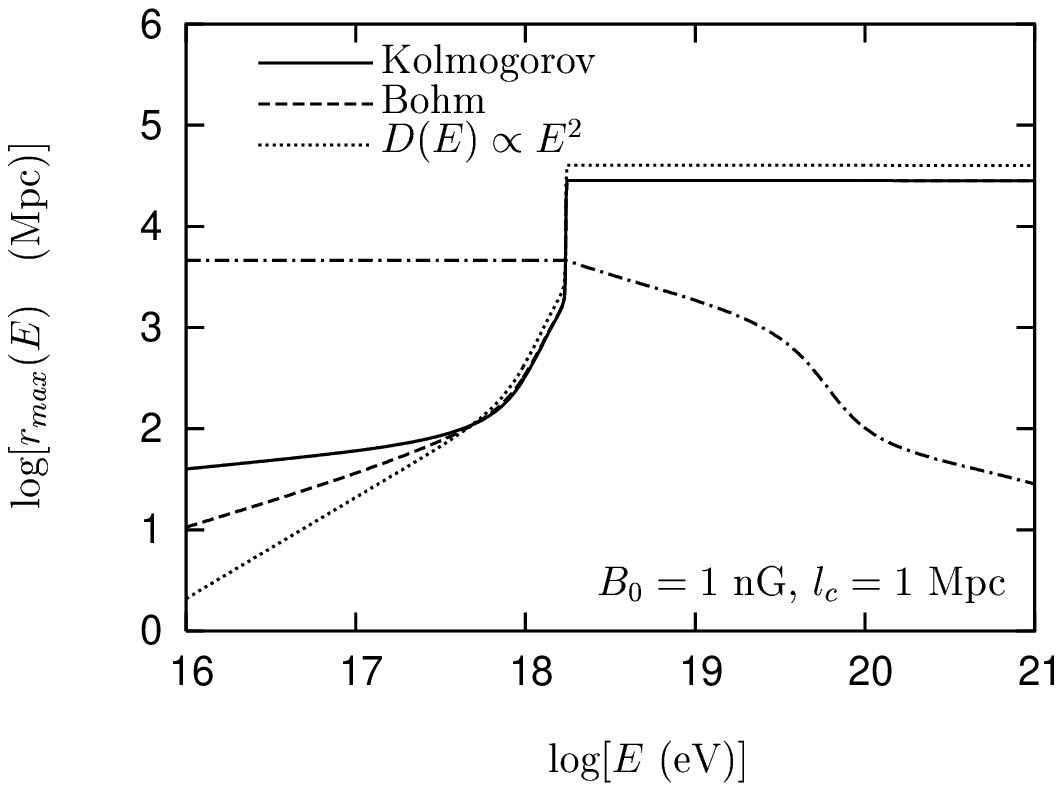}
\\
\end{tabular}
\caption{{\bf [Left Panel]} Maximum generation energy $E_g^{\rm {max}}$ 
defined as ${\rm min}[E_g(E, t_0),~E^{\rm acc}_{\rm max}]$, where 
$E^{\rm acc}_{\rm max}$ is the maximal
acceleration energy and $t_0$ is the age of the universe (see text).
{\bf [Right Panel]} Maximal distance $r_{\rm max}(E)$  to the 
contributing sources as function of the observed energy $E$. Three 
merging curves in the left-low corner give $r^{\rm diff}_{\rm max}(E)$ 
and the dash-dotted curve gives $r^{\rm rect}_{\rm max}(E)$, which 
numerically is very close to the energy-attenuation length 
$l_{\rm att}(E)=[(1/cE)dE/dt)]^{-1}$.}
\label{fig1-2}
\end{figure}

The presence of an IMF is still an open question, the most reliable 
observations of this field are based on the measurement of the Faraday 
rotation (RM) of polarized radio emission \cite{Kronberg}. The upper limit 
obtained with these measurements is $RM<5$ rad/m$^2$, it implies an upper 
limit on the IMF that depends on the assumed scale of coherence length. 
For instance, according to \cite{Blasi99}, in the case of an inhomogeneous 
universe $B_{l_c}<4$ nG with a scale of coherence of about $l_c=50$ Mpc. 
In general, as follows from the observations of Faraday rotation, the magnetic 
field is high, of the order of $1$ $\mu$G with a coherence length $l_c=1$ Mpc, 
in clusters of galaxies and radio lobes of radio galaxies \cite{Kronberg}. 
Apart from observations, the IMF can be predicted, in principle, trough 
Magnetohydrodynamics (MHD) simulations. The main ambiguities in these 
simulations are related to the assumed seed magnetic field and to the 
capability of simulations to reconstruct the local Universe as we 
observe it (i.e. constrained \cite{Dolag} and unconstrained simulations 
\cite{Sigl}). Unfortunately, because of these uncertainties, MHD simulations 
are not completely  conclusive, there are at least two opposite results in 
literature with predicted magnetic field in voids (filaments) that vary 
from $10^{-3}$ nG ($10^{-1}$ nG) \cite{Dolag} up to $10^{-1}$ nG ($10$ nG) 
\cite{Sigl}.

While a direct evaluation of the IMF strength is still challenging,
indirect informations about the UHECR propagation mode can be inferred from 
UHECR data. The analysis of the arrival directions of UHECR at energies 
$E>10^{19}$ eV shows a small angle clustering within the angular 
resolution of the detectors. The AGASA detector has found $3$ doublets 
and $1$ triplet among $47$ detected events \cite{Takeda}. 
This analysis is confirmed also by the combined data of different detectors 
\cite{Takeda} in which $8$ doublets and $2$ triplets are found in $92$
collected events. This evidence can be well understood in terms of a
rectilinear propagation of protons at the highest energies ($E>10^{19}$ eV) 
with a random arrival of two (three) particles from the same source
and a source number density of about $n_s\simeq 10^{-5}$ Mpc$^{-3}$
\cite{BlasiDeM}. However, the small angle clustering may survive in the case
of UHE protons propagation in IMF \cite{Sigl,Sato}.

Another remarkable evidence of an almost rectilinear propagation of UHECR,
in the energy range $2-8 \times 10^{19}$ eV, has been found by 
Tinyakov and Tkachev \cite{TT}. These authors have found a correlation 
between arrival directions of UHE particles in the AGASA and Yakutsk 
detectors and the directions of several BL-Lac objects, (i.e. AGNs 
with jet directed toward us). The combined evidences of small angle 
clustering and correlation with BL-Lacs favor a scenario with a 
quasi-rectilinear propagation of protons at energies larger than $10^{19}$ eV. 

\section{UHECR Diffusive Propagation}
\label{diffusive}

In order to describe the diffusive propagation in IMF of UHECR protons 
we will use the Syrovatsky \cite{Syrovatsky} solution to the diffusive 
equation. Following \cite{anti-GZK}, we will also assume a distribution of 
sources on a lattice; under this hypothesis the diffuse flux can be 
calculated as the sum over the fluxes from the discrete sources at distances 
$r_i$:

\begin{equation}
J_p^{diff}(E)=\frac{c}{4\pi}
\frac{L_p K(\gamma_g)}{b(E)E_{\rm min}^2} \sum_i
\int_{E}^{E_g^{\rm max}} dE_g
\left (\frac{E_g}{E_{\rm min}} \right)^{-\gamma_g} \times
\label{J-diff}
\end{equation}
$$
\times \frac{exp\left [-\frac{r_i^2}{4\lambda(E,E_g)} \right ]}
{\left ( 4\pi\lambda(E,E_g)\right )^{3/2}}~,
$$
where $b(E)=dE/dt$ is the proton energy losses (we have used $b(E)$ as 
computed in \cite{Berezinsky}) and $Q_{inj}=
(L_pK(\gamma_g)/E_{min}^2) (E_g/E_{min})^{-\gamma_g}$ is the 
particle generation rate per unit energy with $L_p$ the source luminosity 
(here we assume identical sources with the same luminosity) and 
$K(\gamma_g)=\gamma_g-2$ (if $\gamma_g>2$) a normalization constant. The 
function $\lambda(E,E_g)$ is 
\begin{equation}
\lambda(E,E_g)=\int_E^{E_g} d\epsilon \frac{D(\epsilon)}{b(\epsilon)}
\label{lambda}
\end{equation}
with $D(E)$ the diffusion coefficient. The quantity in (\ref{lambda}) 
describes the squared distance traversed by a proton in the direction of 
the observer, while its energy decreases from $E_g$ to $E$. 
The Syrovatsky parameter $\lambda(E,E_g)$ poses a natural cut to the 
flux contributing sources, it is clear from equation (\ref{J-diff}) that a 
source at distance $r>2\sqrt{\lambda(E,E_g)}$ gives a negligible 
(exponentially suppressed) contribution to the flux.

The Syrovatsky solution (\ref{J-diff}) formally includes all propagation
times up to $t\to \infty$ and the generation energy of particle is limited
only by the maximum energy that the source can provide $E_{\rm max}$. 
Nevertheless, the propagation time should be smaller than the age of the 
universe $t_0$, this poses an additional limit to the generation energy 
represented by $E_g(E,t_0)$, which can be computed evolving backward 
in time the proton energy from $E$ at $t=0$ up to $E_g$ at $t=t_0$. 
Therefore, to limit the propagation time, the upper limit of integration 
$E_g^{\rm max}(E)$ in equation (\ref{J-diff}) is fixed at the minimum 
between $E_{\rm max}$ and $E_g(E,t_0)$. It is interesting to 
note that at energies $E<E_{\rm s}=1\times 10^{18}$, where only adiabatic 
energy losses are relevant, the upper limit of integration in (\ref{J-diff})
is fixed by $E_g^{\rm max}(E)=E_g(E,t_0)$ while at higher energies, where 
pair-production and photo-pion-production energy losses become relevant, the 
upper limit of integration is $E_g^{\rm max}=E_{\rm max}$. This 
behavior of $E_g^{\rm max}(E)$ is responsible for the low energy 
steepening of the UHECR diffusive spectrum. In Figure \ref{fig1-2} 
(right panel) we have reported the $E_g^{\rm max}$ as function of the 
observed energy $E$, with $E_{\rm max}=1\times 10^{22}$ eV.

\begin{figure}[t!]
\begin{tabular}{ll}
\includegraphics[width=0.45\textwidth]{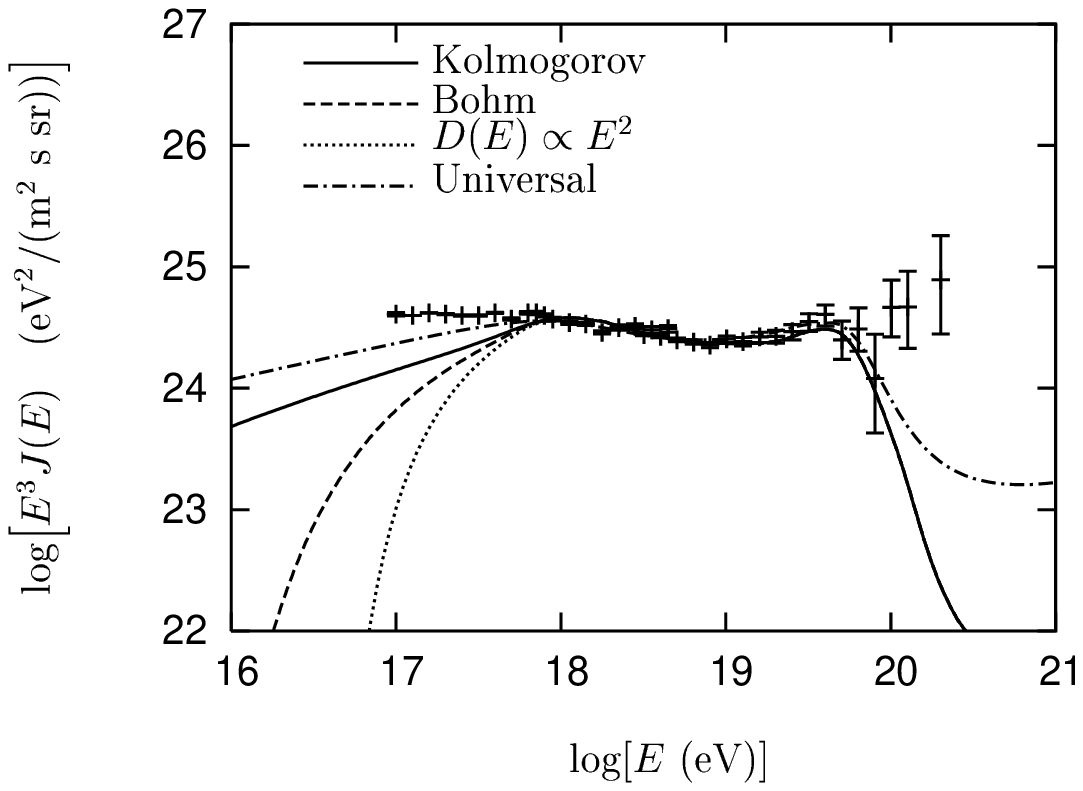}
&
\includegraphics[width=0.45\textwidth]{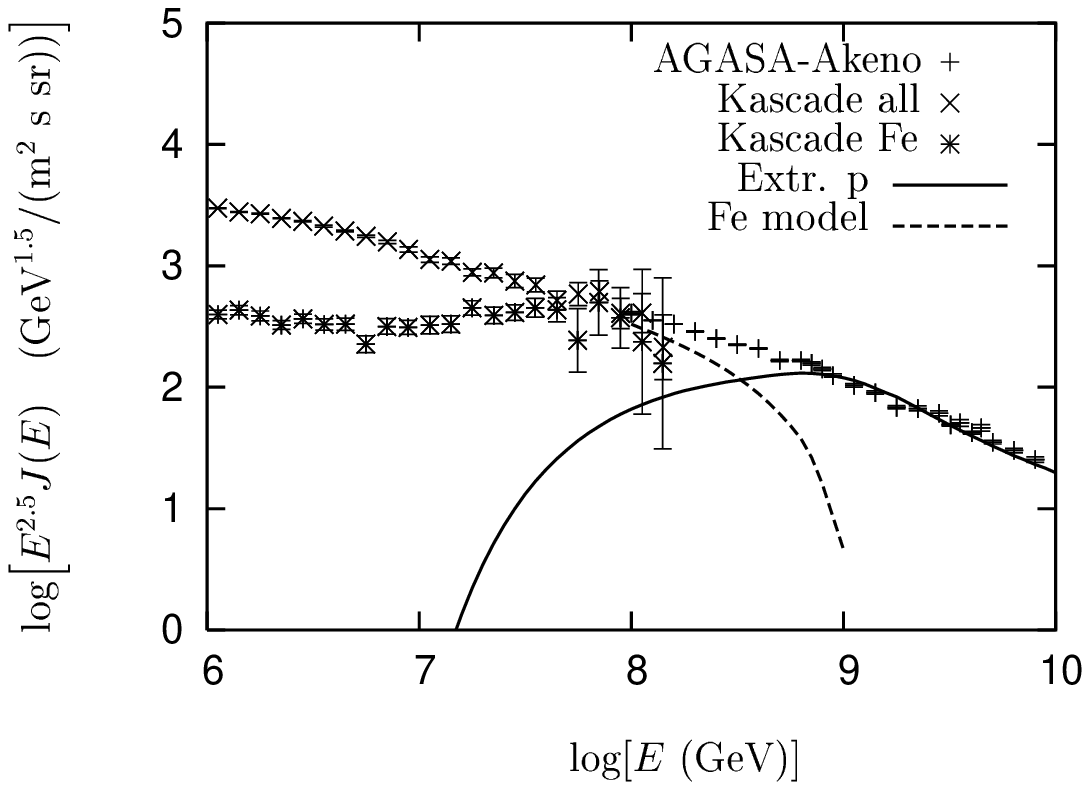}
\\
\end{tabular}
\caption{
{\bf [Left Panel]} Energy spectrum in the case of $B_0=1$ nG, $l_c=1$ Mpc and 
for the diffusion regimes: Kolmogorov (continuous line), Bohm (dashed line) and
$D(E)\propto E^2$ (dotted line). The separation between sources is $d=30$ Mpc 
and the injection spectrum index is $\gamma_g=2.7$ (see text). The AGASA-Akeno 
data with the universal spectrum (dash-dotted line) are also 
reported. {\bf [Right Panel]} The galactic iron-nuclei spectrum computed by 
subtracting the extragalactic proton spectrum from the Akeno-AGASA data. The
extragalactic proton spectrum is taken for the case $B_0=1$ nG, 
$l_c=1$ Mpc, $d=30$ Mpc, $\gamma_g=2.7$ with the Bohm diffusion
at $E < E_c$.
}
\label{fig3-4}
\end{figure}

Let us now concentrate on the diffusion coefficient that enters
in $\lambda(E,E_g)$ function, see equation (\ref{lambda}). Following
\cite{anti-GZK}, we assume diffusion in a random magnetic field with a strength
$B_0$ on the coherence length $l_c$. This assumption determines the diffusion
coefficient $D(E)$ at the highest energies when the Larmor radius of protons 
$r_{L}(E) = 1 (B_0/{\rm nG})^{-1} (E/10^{18} {\rm eV})$ Mpc becomes larger
than $l_c$, namely $D(E)=D_0 (r_L(E)/l_c)^2$ with $D_0=cl_c/3$. 
At low energy (i.e. when $r_L(E)\lsim l_c$) we have 
considered three different cases: 
(i) the Kolmogorov diffusion $D_K(E)=D_0 (r_L(E)/l_c)^{1/3}$; 
(ii) the Bohm diffusion $D_B(E)=D_0 (r_L(E)/l_c)$;
(iii) an arbitrary case $D(E)\propto E^{\alpha}$ with $\alpha=2$ as the
extreme possibility, also in this case $D(E)$ is normalized by $D_0$ at 
$r_L=l_c$, that corresponds to an energy 
$E_c\simeq 10^{18} (B_0/nG) (l_c/Mpc)$ eV.
The diffusion length can be evaluated through the interpolation formula 
$l_{diff}=\Lambda_d + r_L^2/l_c$, with: $\Lambda_d=r_L$ and 
$\Lambda_d=l_c(r_L/l_c)^{1/3}$ in the Bohm and Kolmogorov cases respectively.
At distances $r<l_{diff}(E)$ the proton propagation becomes rectilinear
and the fluxes from individual sources on the lattice are computed in 
the rectilinear propagation limit. In this limit, taking into account
the cosmological evolution of the Universe, the diffuse flux in our 
lattice model will be:

\begin{equation}
J_p^{\rm rect}(E)=\frac{L_p K(\gamma_g)}{(4\pi E_{\rm min})^2} 
\sum_i \left ( \frac{E_g(E,z_i)}{E_{\rm min}} \right )^{-\gamma_g} \times
\label{J-rect}
\end{equation}
$$
\times \frac{1}{r_i^2(1+z_i)} 
\frac{d E_g(E,z_i)}{d E}
$$
where $z_i$ is the red shift that, according to the standard cosmology,
is associated to the source at distance $r_i$, the two quantities 
$E_g(E,z_i)$ and $d E_g(E,z_i)/d E$ can be computed according to 
\cite{Berezinsky}.
 
Following \cite{anti-GZK} we fix a reasonable strength of the IMF: 
namely $B_0=1$ nG on the coherence scale $l_c=1$ Mpc. This 
choice of $(B_0,l_c)$ is compatible with the limits that follows from 
the Faraday rotation measures and with a rectilinear propagation regime at 
the highest energies $E>10^{19}$ eV. In order to reproduce, in our 
lattice model, the density of sources as follows from small angle 
clustering, we have chosen a separation between sources, i.e. a lattice 
spacing, of $d=30$ Mpc that corresponds to a source space density of 
$n_s=1/d^3=3.7\times 10^{-5}$ Mpc$^{-3}$. In the computation of the fluxes we 
have assumed an injection spectrum with a single power law $\gamma_g=2.7$,
starting from the minimum energy $E_{\rm min}=1$ GeV. 

As follows from equation (\ref{J-diff}) the particles that, as a whole, 
contribute to the diffusive flux are those particles produced inside a 
sphere of radius $r_{\rm max}(E)=2\sqrt{\lambda(E,E_g^{\rm max})}$, the 
contribution to the flux is exponentially suppressed for particles produced 
at higher distances. In Figure \ref{fig1-2} (left panel) we have reported, 
fixing $l_c=1$ Mpc and $B_0=1$ nG the behavior of the maximum contributing 
distance $r_{\rm max}(E)$ in the three cases Kolmogorov, Bohm, and 
$D(E)\propto E^2$. The dot-dashed line represents the proton attenuation 
length $l_{\rm att}=E(dE/dl)^{-1}$, that can be interpreted as the maximum 
contributing distance in the rectilinear propagation regime. 
Figure \ref{fig1-2} (right panel) clearly illustrates the steepening of 
the diffusive spectrum at low energies. While the energy-attenuation length 
$l_{\rm att}$ diminishes with energy and has the GZK steepening at energy
$E\simeq 5\times 10^{19}$ eV, the diffusive maximum distance $r_{\rm max}(E)$ 
increases with energy and has a sharp jump at energy 
$E_{\rm eq}\simeq 2\times 10^{18}$ eV. The position of this jump is related 
only to the proton energy losses, it does not depend on the magnetic field 
configuration (i.e. on the diffusion parameters). The energy $E_{\rm eq}$
corresponds to the proton energy at which pair-production energy losses
become equal to adiabatic energy losses. At this energy the upper limit 
of integration in the Syrovatsky solution changes abruptly from $E_g(E,t_0)$ 
to the maximum energy $E_{\rm max}$ that the source can provide. 

Let us consider now the UHECR spectra shown in Figure \ref{fig3-4} 
(left panel). In this figure we report the diffusive and universal spectra,
the latter corresponds to the rectilinear propagation (see equation 
(\ref{J-rect})) in the case of a homogeneous distribution of sources
(see \cite{theorem} for a detailed discussion).
The effect of the low energy steepening is clearly seen in the spectra, with 
the steepening energy $E_s$ being independent of the diffusive regime chosen. 
According to the results presented in Figure \ref{fig1-2} (right panel) 
for $r_{\rm max}(E)$ the flux below the steepening energy $E_s$ is largest 
for the Kolmogorov diffusion and lowest in the case $D(E)\propto E^2$, with 
the Bohm diffusion between them. The source luminosity $L_p$ needed to provide 
the observed spectrum is very high, for a distance between sources of 
$d=30$ Mpc the required luminosity is $L_p=3.0\times 10^{48}$ erg/s. To reduce 
the required emissivity one can assume that the acceleration mechanism works 
only starting from a somewhat higher minimum energy $E_{\rm min}$ 
(see \cite{anti-GZK,model} for a detailed discussion). 

It is also worthwhile to stress that the feature of the dip, that signals
in the experimental data for a proton dominated spectrum, is not washed out by 
the IMF in the diffusive approximation. The validity of this approximation
is related to the validity of the Syrovatsky solution at energies $E<10^{19}$ 
eV. The diffusive equation, and hence its solution, are valid only in the case 
in which the energy losses and diffusion coefficient are time independent. 
At energies lower than $10^{19}$ eV, this is not the case, because the 
propagation time of protons approaches the age of the universe and the effect 
of the CMB temperature variation with time (red-shift) becomes important. 
However, our computations show a good agreement between the quasi-rectilinear 
regime of propagation and the exact rectilinear propagation, this agreement 
shows the approximate validity of the Syrovatsky solution at the discussed 
energies.

Following \cite{Lemoine,anti-GZK,BGH} we shall conclude by discussing shortly 
the transition from galactic to extragalactic cosmic rays. The remarkable 
feature of the diffusive spectra is the low-energy steepening at the fixed 
energy $E_s \sim 1\times 10^{18}$~eV, which provide the transition from
extragalactic to galactic CR. This energy coincides approximately with the
position of the 2nd knee $E_{\rm 2K}$ and gives a non-trivial explanation
of its value as $E_{\rm 2K} \sim E_{s}$. Like in the above-mentioned works we 
shall assume that at $E \gsim 1\times 10^{17}$~eV the galactic spectrum is 
dominated by iron nuclei and calculate their flux by subtracting the 
calculated flux of extragalactic protons from all-particle Akeno spectrum. 
For these calculations we shall fix the spectrum with magnetic configuration 
(1~nG, 1~Mpc), the Bohm diffusion and a separation between sources on the 
lattice $d=30$~Mpc The calculated spectrum of galactic iron is shown in 
Figure \ref{fig3-4} (right panel) by the dashed curve. This prediction should 
be taken with caution because obtained with a model-dependent calculations 
(assumption of the Bohm diffusion) and uncertainties involved in the 
Syrovatsky solution. However, it is interesting to note that the iron-nuclei 
spectrum in Figure \ref{fig3-4} (right panel) is well described by the Hall 
diffusion \cite{Ptuskin} in the galactic magnetic field at energies above 
the knee.

\section{Conclusions}
\label{conclusions}

We have analyzed the anti-GZK effect in the diffusive propagation of UHE 
protons. This effect consists in a sharp increase of the maximum distance 
$r_{\rm max}(E)$ from which UHE protons can arrive (Figure \ref{fig1-2} 
(left panel)). The observational consequences of the anti-GZK effect is a 
low-energy steepening of the diffuse spectrum. While the shape of the 
steepening depends on the magnetic field configuration, the steepening energy 
$E_s$ is practically independent of it. The steepening of the spectrum at 
$E_s \sim 1\times 10^{18}$~eV coincides with the 2nd knee observed in the CR 
spectra by most of the detectors and provides a natural transition from 
galactic iron nuclei to extragalactic protons.

\end{document}